\documentclass[aps,prb,twocolumn,groupedaddress,superscriptaddress,showpacs,floatfix]{revtex4-1}
\usepackage{graphicx}

\usepackage{xcolor}
\usepackage{amsfonts}
\usepackage{amsmath}
\usepackage{amssymb}

\usepackage[english]{babel}

\begin{document}

\title{Topological Phases in Triangular Lattices of Ru Adsorbed on 
Graphene: \textit{ab-initio} calculations}

\author{C. Mera Acosta}
\email[]{acosta@if.usp.br}
\affiliation{Instituto de F\'isica, Universidade de S\~ao Paulo, CP 66318, 05315-970, S\~ao Paulo, SP, Brazil}

\author{Matheus P. Lima} 
\email[]{mplima@if.usp.br}
\affiliation{Instituto de F\'isica, Universidade de S\~ao Paulo, CP 66318, 05315-970, S\~ao Paulo, SP, Brazil}

\author{R. H. Miwa} 
\email[]{hiroki@infis.ufu.br}
\affiliation{Instituto de F\'isica, Universidade Federal de Uberlandia, Uberlandia, MG, Brazil}

\author{Ant\^onio J. R. da Silva} 
\email[]{jose.roque@lnls.br}
\affiliation{Instituto de F\'isica, Universidade de S\~ao Paulo, CP 66318, 05315-970, S\~ao Paulo, SP, Brazil}
\affiliation{Laborat\'orio Nacional de Luz S\'{\i}ncrotron, CP 6192, 13083-970, Campinas, SP, Brazil}

\author{A. Fazzio} 
\email[]{fazzio@if.usp.br}
\affiliation{Instituto de F\'isica, Universidade de S\~ao Paulo, CP 66318, 05315-970, S\~ao Paulo, SP, Brazil}


\begin{abstract}

We have performed an {\it ab initio} investigation of the electronic properties
of the graphene sheet adsorbed by Ru adatoms (Ru/graphene).  For a particular
set of triangular arrays of Ru adatoms, we find the formation of four
(spin-polarized) Dirac cones attributed to a suitable overlap between two
hexagonal lattices: one composed by the C sites of the graphene sheet, and the
other formed by the surface potential induced by the  Ru adatoms.  Upon the
presence of spin-orbit coupling (SOC) nontrivial band gaps take place at the
Dirac cones promoting several topological phases. 
Depending on the Ru concentration, the system can be topologically characterized
among the phases i) Quantum Spin Hall (QSH), ii) Quantum Anomalous Hall(QAH), iii) metal
iv) or trivial insulator. For each concentration, the topological phase is characterized 
by the ab-initio calculation of the Chern number. 

\end{abstract}

\pacs{81.05.ue 73.43.Lp 31.15.A-}
\maketitle

\section{Introduction}

The electronic properties of graphene can be tailored in a suitable way through
the deposition of foreign atoms, allowing not only the development of
electronic devices but also providing platforms for new physical phenomena~\cite{CastroNeto20091094,
PhysRevB.82.033414, PhysRevB.81.115427}.
In particular, the adsorption of transition metals (TMs) on the graphene sheet
(metal/graphene) has been considered as a promising  route to modify the
electronic properties of graphene. For instance, the control of spin-polarized
current in graphene by the deposition of TMs (Mn, Fe Co and Ni)~\cite{Lima,PhysRevB.84.235110}. 

Recently, it was shown that it is possible to increase the Spin-Orbit Coupling
(SOC) of graphene by depositing heavy atoms, such as indium and
thallium~\cite{Weeks2011}. In this case, the absence of  net (local) magnetic
moment preserves the time-reversal (TR) symmetry, giving rise to a Quantum
Spin-Hall (QSH) state~\cite{Kane2004} on the metal/graphene sheet with a
nontrivial bulk band gap about three orders of magnitude greater than the
predicted gap in pristine graphene. On the other hand, the majority of TMs with partially
filled $d$ orbitals, adsorbed on the graphene sheet, may promote a local net
magnetic moment, and thus, will suppress the QSH state. However, based on the {\it
ab initio} calculations, Hu {\it et al.} verified that such magnetic moment can
be quenched by either applying an external electric field, or by a codoping processes,
thus, recovering the QSH state~\cite{Hu2012}. They have considered ($n\times
n$) periodic as well as random distributions of TM adatoms on the graphene sheet,
and in both cases the QSH state was preserved. On the other hand, the
appearance of a nontrivial energy bandgap in metal/graphene systems, due to 
the SOC and nonzero magnetic moment (breaking the TR symmetry), gives rise to
the so called  Quantum Anomalous Hall (QAH)
effect~\cite{Qiao2010,Ding2011,Qiao2012,chang2013experimental}. Furthermore, in
a random distribution of TM adatoms on graphene sheet the SOC is not affected,
and the intervalley $K$ and $K'$ scattering is somewhat
suppressed\cite{PhysRevLett.109.116803}. That is, the nontrivial topological
phase of metal/graphene was preserved. Moreover, the tunning process of
QAH effect, by application of external electric field, has been
proposed~\cite{Zhang2012a} for metal/graphene systems adsorbed with 5$d$ TMs.

Those findings allow us to infer that the electronic properties as well as the
topological phases of metal/graphene systems can be controlled  by means of a
geometrical and chemical manipulation, as well as by application of external fields. In a
recent experiment, Gomes {\it et al.}~\cite{Gomes2012} showed that it is possible
to build up artificial graphene structures, so called ``molecular graphene'', on
solid surfaces by the manipulation of the surface potential. They have
considered CO molecules forming a triangular lattice over the Cu(111) surface,
CO/Cu(111). Such surface engineering allows a number of degrees of freedom to
create of artificial lattices that exhibit a set of desirable electronic
properties. For instance, the electronic properties of such ``molecular
graphene'' can be tunned by changing the lateral distance between the CO
molecules, by choosing another molecules instead of CO, or another surface instead
of Cu(111).  Indeed, recent works discuss the possibility of tuning the electronic
properties of molecular graphene, as well as the realization of QSH state in  CO/Cu(111)
surfaces~\cite{Polini,PhysRevB.86.201406}. 
Thus, it is experimentally possible to manipulate atoms and molecules in order
to form an ordered array on top of a substrate.  In the same sense, the usage of
graphene as a substrate for adatoms or foreign molecules may also be
interesting. In this case, the adatoms or molecules will be embedded in a two
dimensional electron gas formed by the $\pi$ orbitals of graphene. 

In this work we performed an investigation of the interplay between
the electronic properties and the geometry of Ru arrays adsorbed on the graphene
sheet (Ru/graphene). We show that by changing the concentration of Ru adatoms
it is possible to cover multiple topological phases\cite{oh2013complete}. 
Thus, with the same transition metal atom and the same triangular lattice 
structure, one has in the lattice parameter (or TM separation) of this 
superstructure a dial that allows to tune the topological properties of the material.

The Ru adatoms, independently of their lattice geometry, strongly interact with
graphene and locally modify the charge density at their neighboring carbon
atoms. Due to the
Ru$\leftrightarrow$Ru indirect interaction via graphene, the electronic structure of Ru/graphene
system becomes ruled by the Ru lattice geometry. We have considered Ru
adatoms forming triangular lattices with ($n \times n$) periodicities, with
respect to the graphene unitary cell, and according to the electronic structure
we found three typical families of periodicities, showed in the Fig. \ref{Fig1}.
For ($3n\times 3n$) periodicity, the Dirac cones are
suppressed due to intervalley ($K$ and $K'$) scattering process, leading to a 
trivial bandgap. Whereas, for the both $((3n+1)\times (3n+1))$, and
$((3n+2)\times(3n+2))$ there is a multiplicity of Dirac cones, and the
appearance of a QAH phase. 
For these two last families, to better understand how the QAH phase emerges, we
sequentially include the effects of an (i) electrostatic potential , (ii) an exchange
potential, and (iii) the spin-orbit coupling. Considering only the electrostatic
potential, two spin degenerated Dirac cones occur (at $K$ and $K^\prime$) due to the
presence of two overlapping hexagonal lattices, one composed by the C atoms of
the graphene sheet, and the other composed by the surface potential on the
graphene sheet induced (lying) by (on the barycenter of) the triangular lattice
of the Ru adatoms. Upon inclusion of the exchange field, due to the net magnetic
moment of Ru adatoms, there is a spin-splitting of all Dirac cones, leading to
an amazing crossing between bands with reverse spin very close to the Fermi level. And
finally, by turning on the SOC, the Rashba Spin-Orbit interaction couples the
reverse-spin states leading to a non-trivial bandgap opening around the Fermi-level.
Thus, the sum of the electrostatic potential, exchange field, and the SOC coupling
leads to non-trivial topological phases in Ru/graphene systems. 

We also show that the topological phase of the Ru/graphene systems changes for higher 
concentrations of Ru adatoms. Ru/graphene with the $(2\times2)$ periodicity presents a 
QSH phase, while for the $(4\times4)$ periodicity the system presents a metalic
phase. The topological classification of the studied systems was made by the
{\it ab-initio} calculation of the Chern number.

\section{Methodology}

All results presented in this work were obtained with first-principles
calculations performed within the Density Functional Theory (DFT)
framework\cite{capelle2006bird}, as implemented in the SIESTA
code\cite{soler2002siesta}. The Local Density Approximation
(LDA)\cite{perdew1981self} is used for the exchange-correlation functional. We
used an energy cutoff of 410 Ry to define the grid in the real space, and the
supercell approximation with a k-points sampling for the reciprocal space
integration equivalent to $20\times20\times1$ in the unitary cell. The 2D
graphene sheets lie in the xy plane, and a vacuum of 20\AA\ was used in the
z-direction to avoid the undesirable interaction between the periodic images of
graphene sheets. All the configurations of the Ru/graphene systems were fully
relaxed until the residual forces on the atoms were smaller than 0.01 eV/\AA. 

In order to investigate the non-trivial topological phases in the Ru/graphene
systems, we implemented the Spin-Orbit Coupling in the SIESTA code within the
on-site approximation\cite{fernandez2006site}.
Within this approach, the Kohn-Sham Hamiltonian $\boldsymbol{H}$
is a sum of the  kinetic
energy $\boldsymbol{T}$, the Hartree potential $\boldsymbol{V}^{H}$,
the exchange and correlation potential $\boldsymbol{V}^{xc}$,
the scalar relativistic ionic pseudopotential $\boldsymbol{V}^{sc}$, and the
spin-orbit interaction $\boldsymbol{V}^{SOC}$. $\boldsymbol{H}$ can be written
as a $2\times2$ matrix in the spin space as:
\begin{equation}
\boldsymbol{H}=\boldsymbol{T}+\boldsymbol{V}^{H}+\boldsymbol{V}^{xc}+\boldsymbol{V}^{sc}+\boldsymbol{V}^{SOC}
=\left[\begin{array}{cc}\boldsymbol{H}^{\uparrow\uparrow} & \boldsymbol{H}^{\uparrow\downarrow}\\
\boldsymbol{H}^{\downarrow\uparrow} & \boldsymbol{H}^{\downarrow\downarrow}\end{array}\right].
\end{equation} 
All terms contribute to the diagonal elements, however only the
$\boldsymbol{V}^{xc}$ and the $\boldsymbol{V}^{SOC}$ potentials have
off-diagonal coupling terms due to the non-collinear spin. 
The spin-orbit matrix elements, as implemented in this work, are written as: 
\begin{equation}\label{eqls1}
V_{ij}^{SOC}=\frac{1}{2}V^{SOC}_{l_{i},n_{i},n_{j}}
\langle l_{i},M_{i}|\boldsymbol{L}\cdot \boldsymbol{S}|l_{i},M_{j}\rangle\delta_{l_{i}l_{j}}, 
\end{equation}
where $|l_{i},M_{j}\rangle$ are the real spherical harmonics\cite{Blanco199719}. 
The radial contributions 
$V^{SOC}_{l_{i},n_{i},n_{j}}= \langle
R_{n_{i},l_{i}}|V_{l_{i}}^{SOC}|R_{n_{j},l_{i}}\rangle$
are calculated with the solution of the Dirac equation for each atom.
The angular contribution $\boldsymbol{L}\cdot \boldsymbol{S}$, considering
the spin operator in terms of the Pauli matrices, can be written as:
\begin{equation}\label{eqls2}
\boldsymbol{L}\cdot \boldsymbol{S} 
=\left[\begin{array}{cc} {L}_{z} &  {L}_{-}\\
 {L}_{+} & -{L}_{z}.\end{array}\right].
\end{equation}
The diagonal matrix elements for the SOC term $V^{SOC,\sigma\sigma}_{ij}$
(with $\sigma=\uparrow$ or $\downarrow$),
are proportional to $\langle l_{i},M_{i}|L_{z}|l_{i},M_{j}\rangle$, which are different from zero only for  $M_{i}=\pm M_{j}$. 
Thus, these terms couple orbitals with the same spins, and same $|M|$.
On the other hand, the off-diagonal matrix elements $V^{SOC,\sigma-\sigma}_{ij}$ are
proportional to $\langle l_{i},M_{i}|L_{\pm}|l_{i},M_{j}\rangle$, and thus
couple orbitals with different spins and $M_{i}=M_{j}\pm1$. These coupling terms
could open bandgaps or generate the inversion of states that are essential to
the physics of the topological insulators. 

The band gaps were topologically characterized by calculating the
Chern number ($\mathcal{C}$). This number is necessary to identify the topological
class induced by the SOC in magnetic systems and is related 
to non-trivial Hall conductivity.
In two dimensional systems the Chern number can be calculated within a
non-Abelian formulation\cite{PhysRevB.85.115415} by the following expression:
\begin{equation}\label{eq1}
 \mathcal{C}=\frac{1}{2\pi}\int_{BZ}\text{Tr}[\boldsymbol{B}(\boldsymbol{k})]d^2k.
\end{equation}
Where the trace is a summation over the band index, and only the occupied bands
are taken in account.
The integration is done over the whole Brillouin Zone (BZ), and
$\boldsymbol{B}(\boldsymbol{k})$
is a matrix representing the non-abelian momentum-space Berry curvature, whose
diagonal elements can be written as \cite{PhysRevB.85.115415}:
$$\boldsymbol{B}_{n}(\boldsymbol{k})=\underset{\Delta_{k_y}\rightarrow0}{\text{lim}}\underset{\Delta_{k_x}\rightarrow0}{\text{lim}}\frac{-i}{\Delta_{k_x}\Delta_{k_y}}
\text{Im}\text{ log}[\langle u_{n\boldsymbol{k}}\rvert
u_{n\boldsymbol{k}+\Delta_{k_x}}
\rangle\times$$
$$\langle u_{n\boldsymbol{k}+\Delta_{k_x}}\rvert
u_{n\boldsymbol{k}+\Delta_{k_x}+\Delta_{k_y}}\rangle
\langle u_{n\boldsymbol{k}+\Delta_{k_x}+\Delta_{k_y}}\rvert
u_{n\boldsymbol{k}+\Delta_{k_y}}\rangle 
\times$$
\begin{equation}
\langle u_{n\boldsymbol{k}+\Delta_{k_y}}\rvert u_{n\boldsymbol{k}}\rangle].
\end{equation}
Where $\Delta_{k_x}$ ($\Delta_{k_y}$) is the grid displacement in the $k_x$
($k_y$) direction of the reciprocal space,   $\rvert
u_{n\boldsymbol{k}}\rangle$ is the cell-periodic Bloch functions in the
($\boldsymbol{k}$) point of the BZ, and $n$ indicates the band index. 
This expression is quite adequate to perform calculations in systems  with 
band crossing, and was implemented using a discrete grid in the reciprocal space.

\section{Results}
\begin{figure*}
\includegraphics[width = 18cm]{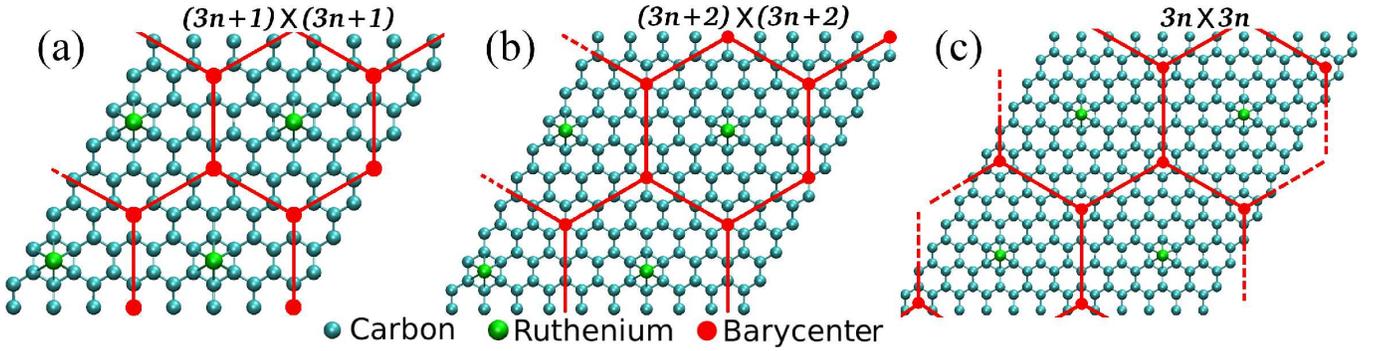}
\caption{(Color online) Schematic representations of Ru/graphene systems with
(a) ($4\times4$), (b) ($5\times5$) and (c) ($6\times6$)  periodicities, respectively. 
These are examples of the $((3n+1)\times(3n+1))$, $((3n+2)\times(3n+2))$ and
($3n\times3n$) Ru/graphene systems, respectively. Here, $n$ is an integer. The 
geometric centers of the triangles formed by the Ru adatoms (barycenters) 
form a honeycomb lattice, represented in the figure.}
\label{Fig1}
\end{figure*} 

The energetic stability of Ru adatoms on the graphene sheet was examined through
the calculation of the binding energy ($E^b$) written as, 
$$
E^{b}=E[\mbox{graphene}]+E[\mbox{adatom}]-E[\mbox{Ru/graphene}]. 
$$ 
$E[\mbox{graphene}]$ and $E[\mbox{adatom}]$ represent the total energies of the
separated systems, graphene sheet and an isolated Ru atom, respectively, and
$E[\mbox{Ru/graphene}]$ represents the total energy of the (final) Ru adsorbed
graphene sheet, Ru/graphene. We have considered Ru/graphene systems with
($n \times n$) periodicity with $n$ ranging from 2 up to 12, thus, a set of
different Ru concentrations on the graphene sheet. For the energetically most
stable configuration, the Ru adatom presents a $C_{6v}$ symmetry,  sitting on
the hollow site ($H$) of the graphene sheet. For the ($4\times 4$) periodicity,
we obtained  $E^b = 2.64$~eV at $H$, while  the top site (above the C atom) is
energetically less stable by 0.73~eV ($E^b = 1.91$~eV). There is a negligible
dependence between the calculated binding energies and the Ru concentration. We
did not find any energetically stable configuration for Ru adatoms on the 
bridge site (on the C--C bond).  It is noticeable that the Ru binding energy is
larger when compared with most of transition metals (TMs) adsorbed on
graphene\cite{Ding2011,PhysRevB.77.195434,PhysRevB.77.235430}. At the
equilibrium geometry the Ru adatom lies at 1.68~\AA\ from the graphene sheet
(vertical distance $z$), which is smaller when compared with most of the other
TMs on graphene\cite{PhysRevB.77.235430}. Those findings allow us to infer that
there is a strong chemical interaction between Ru adatoms and the graphene
sheet. Indeed, our electronic structure calculations indicate that the Ru-4$d$
orbitals, $d_{x^{2}-y^{2}}$, $d_{xz}$, $d_{yz}$ and $d_{yx}$, are strongly
hybridized with the  carbon $\pi$ (host) orbitals, while $d_{z^2}$ 
behaves as lone pair orbitals.

\begin{figure}
\begin{center}
\includegraphics[width = 8.5cm ]{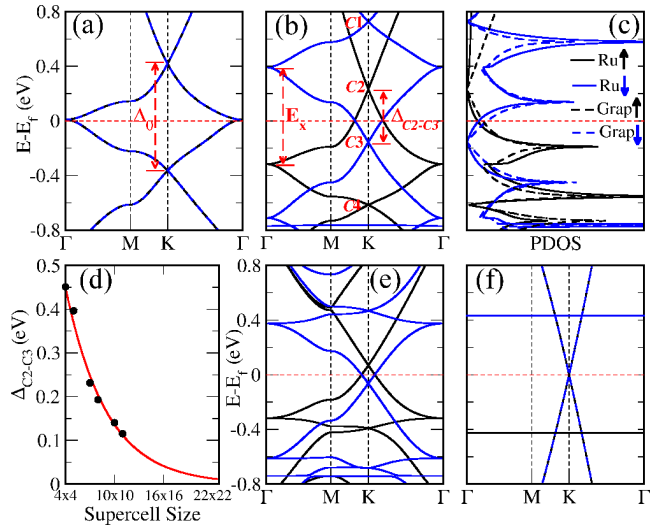}
\end{center}
\caption{(Color online) Evolution of the electronic band structure 
of Ru/graphene with a ($5\times5$) periodicity, considering successively the contributions of 
(a) electrostatic potential generated by Ru adatoms and (b) exchange field.
In (c) is shown the PDOS of (b).
Dark gray (Blue) and black lines are associated with the down and up spin,
respectively.
(d) Variation of the energetic separation, $\Delta_{C2-C3}$, of the Dirac cones closer to the
Fermi level ($C2$ and $C3$) relative to the concentration of Ru adatoms.
(e) Band structure of (10$\times$10) Ru/grapehene system.
(f) Expected picture of the band structure of ($n \times n$) Ru/graphene systems
with $n\rightarrow\infty$.}
\label{Fig2}
\end{figure}

Initially we will examine the electronic properties of Ru/graphene systems with
$((3n+1)\times(3n+1))$ and $((3n+2)\times(3n+2))$ periodicities, which
geometries are schematically represented in the panels (a) and (b) of Fig.
\ref{Fig1}, respectively. The ($3n\times 3n$) [see Fig.~\ref{Fig1}(c)] systems
will be discussed later on. Due to the Ru induced electrostatic field on the
graphene sheet,  the (5$\times 5$) Ru/graphene system exhibits two
spin-degenerated band intersections (Dirac cones), at the $K$ and $K^\prime$
points, separated by 0.78~eV [indicated as $\Delta_0$ in
Fig,~\ref{Fig2}(a)]. Further inclusion of spin polarization gives rise to a
sequence of four spin-split band intersections, $C1-C4$ in Fig.~\ref{Fig2}(b).
The strength of the exchange field can be measured by the energetic separation
$E_{x}$ at the $\Gamma$ point, of 0.66~eV, as shown in Fig. \ref{Fig2}(b).
In the same diagram,
$\Delta_{C2-C3}$ indicates the energy separation between the highest occupied
($C3$) and lowest unoccupied ($C2$) Dirac cones. Notice that the linear energy
dispersion (Dirac cones) has been preserved. Our calculated Projected Density of
States (PDOS) [Fig.~\ref{Fig2}(c)] reveals that the Dirac cones are composed by
similar contributions from C $2p$ ($\pi$ orbitals) and Ru $4d$ orbitals. On the
other hand, reducing the Ru adatom concentration by increasing the  ($n \times
n$) periodicity, we find that, (i) the electronic contribution of Ru $4d$
orbitals to the Dirac cones $C2$ and $C3$ ($C1$ and $C4$) reduces (increases);
(ii) in contrast, the C $\pi$ orbitals contribution to $C1$ and $C4$ ($C2$ and
$C3$) reduces (increases); (iii) the energy dispersions of the electronic bands
that form the Dirac cones $C1$ and $C4$ have been reduced (becoming flatter). The
localized character of Ru $4d$ orbitals will be strengthened, in accordance with
(i). (iv) The electronic bands $C2$ and $C3$ retrieve the behavior of
pristine graphene sheet, in accordance with (ii). The role played by the
Ru adatom becomes negligible, and   $\Delta_{C2-C3} \rightarrow 0$ for larger
($n \times n$) periodicity, as shown in Fig.~\ref{Fig2}(d). In
Fig.~\ref{Fig2}(e) we present the electronic band structure of (10$\times$10)
Ru/graphene, where (iii) and (iv) described above can be verified, and in
Fig.~\ref{Fig2}(f) we present the expected picture of the electronic band
structure of ($n \times n$) Ru/graphene for $n\rightarrow\infty$. Those
findings confirm the strong electronic coupling between the Ru adatoms and the
graphene sheet (Ru$\leftrightarrow$graphene)  leading to a long range
interaction between the Ru adatoms via graphene (Ru$\leftrightarrow$Ru).

The total magnetic moment per Ru atom is also found to depend on the Ru
coverage. Apart from the $(2\times2)$ periodicy which is non-magnetic, all
studied structures present a finite magnetic moment. For all the $(3n\times3n)$
periodicities the magnetic moment is 2.0$\mu_B$. Whereas, for the
$((3n+1)\times(3n+1))$ and $((3n+2)\times(3n+2))$ periodicities the magnetic
moment increases with $n$ from 1.75 to 2.0$\mu_B$ at the limit of low coverage.

\begin{figure}
\includegraphics[width = 8.5cm ]{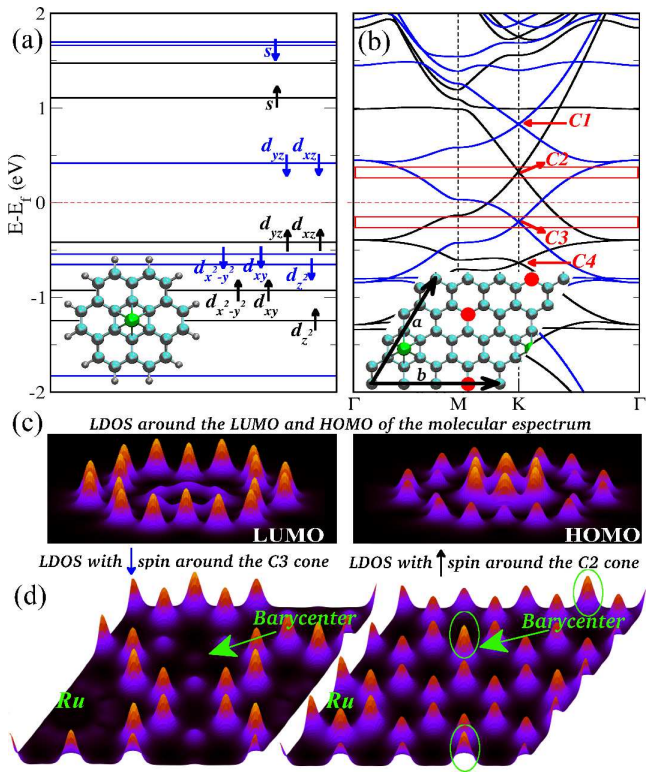}
\caption{(Color online) Local effect of Ru adatom. (a) Molecular spectrum of Ru adsorbed 
on the coronene molecule, as illustrated in the structure at the bottom left
of the box. (b) Electronic band structure of (4$\times$4) Ru/graphene system without
spin orbit coupling. 
The (red) boxes around C2 and C3 indicate the energy intervals used to calculate
the LDOS. 
In the bottom left is represented the structure of the unit cell used.
We indicate the barycenters with (red) balls.
(c) LDOS around the LUMO (left) and HOMO (right) of molecular spectrum in
arbitrary units.  (d) LDOS with down and up spins around the $C3$ and $C2$
cones, respectively.  
We point out a C site associated with a barycenter, which is indicated
in the structure in (b).
}
\label{Fig3}
\end{figure}

In order to improve our understanding on the role played by the (long range)
Ru$\leftrightarrow$Ru electronic interaction, upon the presence of graphene
host, we turned off such Ru$\leftrightarrow$Ru interaction by examining the
electronic structure of a Ru adatom adsorbed on the central hexagon of a
coronene molecule ($C_{24}H_{12}$), Ru/coronene. This geometry is represented in
the inset of Fig.\ref{Fig3}(a). This is a hypothetical system, since the
equilibrium geometry of the Ru adatom in Ru/coronene is kept as the same as that
obtained for the periodic Ru/graphene system. The calculated molecular spectrum,
presented in Fig.~\ref{Fig3}(a), reveals that the HOMO and LUMO are both
bi-degenerated states (mostly) composed by $d_ {xz}$ and $d_{yz}$ orbitals of Ru
adatom, being the spin-up ($\uparrow$) component for the HOMO and spin-down
($\downarrow$) for the  LUMO. The effect of Ru$\leftrightarrow$Ru interaction,
mediated by the graphene $\pi$ orbitals, can be observed by comparing the 
panels (a) and (b) in Fig.~\ref{Fig3}. In Fig.~\ref{Fig3}(b), we present the
electronic band structure of ($4\times4$) Ru/graphene, where it is noticeable
that the HOMO and LUMO energies of Ru/coronene compare very well with those of
($4\times4$) Ru/graphene at the $\Gamma$ point. We also identify the
other $4d$ states ($d_{z^2}$, $d_{x^2-y^2}$ and $d_{xy}$). In the Ru/graphene
system, the Ru $4d$ orbitals (hybridized with $\pi$ orbitals of the graphene
sheet) exhibit a dispersive character along the $\Gamma$--M and $\Gamma$--K
directions within the Brillouin zone.
We find that the Dirac cones with spin-up (spin-down) above (below) the Fermi
level, indicated as $C2$ ($C3$) in Fig.~\ref{Fig3}(b), are formed by
dispersive states with contributions of the $d_{xz}$, $d_{yz}$,
$d_{x^{2}-y^{2}}$, and $d_{xy}$ Ru orbitals. In other words, the states around
the Fermi energy are composed by Ru orbitals with $l=2$ and $m=\pm1,\pm2$, hybridized
with the carbon $p_z$ orbitals.


In Figs.~\ref{Fig3}(c) and \ref{Fig3}(d) we present the  Local Density of States
(LDOS) of the Ru/coronene and (4$\times$4) Ru/graphene systems, respectively. In
those diagrams, the electronic states were projected onto a parallel plane to
the Ru/coronene and Ru/graphene interfaces, 0.5~\AA\ above the molecule and the
graphene sheet, respectively.  We verify that the molecular spectrum of the HOMO
and LUMO [Fig.~\ref{Fig3}(c)] are localized on the nearest neighbor (NN) and the
next nearest neighbor (NNN) C sites of the Ru adatom, respectively. It is
noticeable that for periodic Ru/graphene systems, regardless of the size and
geometry of Ru adatoms, we have the similar electronic distribution for the
occupied and empty states at the $\Gamma$ point. {\it viz.}:  the spin-up
(spin-down) C $\pi$ orbitals, localized on the NN (NNN) neighbor sites of the Ru
adatom, contribute to the formation of the highest occupied (lowest unoccupied) states. 
Thus, we can infer that the Ru adatom locally defines the charge
density both at its first and second nearest neighbor carbon atoms. 
On the other hand, due to the energy dispersion of those states along the
$\Gamma\rightarrow K$ direction, the electronic states of the Dirac cone $C2$
will be mostly localized on the C atoms NN to the Ru adatoms
[Fig.~\ref{Fig3}(d-right)], while the C atoms NNN to the Ru adatom
[Fig.~\ref{Fig3}(d-left)] will contribute to the formation of $C3$. In addition,
 as shown in Fig.~\ref{Fig3}(d), not only the C atoms NN and NNN to the Ru
adatom contribute to the formation of the Dirac cones, but also there are
electronic contributions from the other C atoms of the graphene sheet. 
In particular, the LDOS of $C2$ exhibits a constructive wave function
interference (LDOS peak) on the carbon atom lying at the geometric center
of the triangular array of Ru adatoms, called hereafter as barycenter [indicated
by an arrow in Fig.~\ref{Fig3}(d)], while it becomes destructive for the
Dirac cone $C3$ [Fig.~\ref{Fig3}(d-left)]. 

\begin{figure}
\includegraphics[width = 8.5cm ]{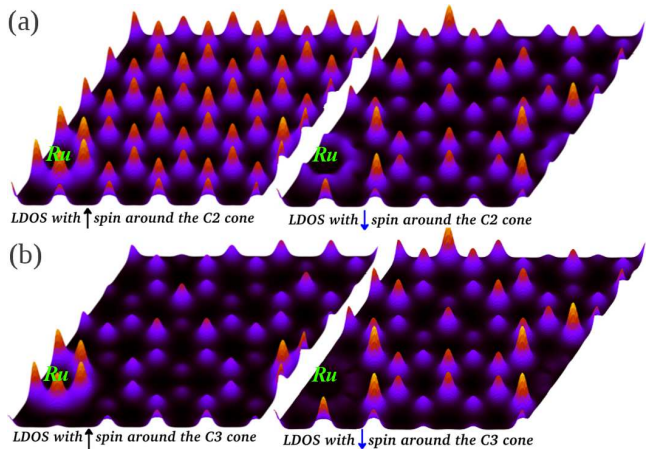}
\caption{(Color online) LDOS for up and down spins around the energy level at
which the (a) $C2$ and (b) $C3$ cones are formed for the (5$\times$5) Ru/graphene system.
}
\label{Fig4}
\end{figure}

Further LDOS calculations reveal that the other Dirac cone  above $E_F$, $C1$
(spin-down), presents similar electronic distribution as compared to $C2$,
whereas the LDOS of the Dirac cone $C4$ (spin-up), below $E_F$, is similar to that of $C3$.
We find the same LDOS picture for the (7$\times$7) and (10$\times$10) Ru/graphene
systems, namely, the Dirac cones above $E_F$ present LDOS peaks (i) on the NN C
atoms to the Ru adatom and (ii) on the C atom localized at the barycenter of
the triangular array of Ru adatoms. While the LDOS of the Dirac cones below
$E_F$ are (iii) localized on the NNN C atoms to the Ru adatoms, and (iv)
present a negligible electronic contribution from the barycenter C atom. 
Such electronic picture, as described in (i)--(iv), is verified 
for the other ($(3n+1)\times (3n+1)$) family of Ru/graphene systems, where the
barycenter and the NN C sites to the Ru adatom belong to the same sublattice.
Meanwhile, for the $((3n+2)\times (3n+2))$ Ru periodicity,
such as  (5$\times$5), (8$\times$8), and (11$\times$11) Ru/graphene systems, the
barycenter and the NNN C sites to the Ru adatom belong to the same sublattice.
Such difference gives rise to a distinct LDOS picture for the Dirac
cones. In Fig.~\ref{Fig4}(a-left) and \ref{Fig4}(b-right) we present the LDOS
for the (5$\times$5) Ru/graphene, for the spin-up Dirac cone $C2$ and spin-down Dirac
cone $C3$, above and below $E_F$, respectively. Here, compared with the
(4$\times$4) counterpart, in the (5$\times$5) Ru/graphene system the Dirac
cones above $E_F$ obey (i) and (iv), whereas the Dirac cones below $E_F$ are
characterized by (ii) and (iii). Thus, we can infer that the electronic states
at the barycenter C atom contributes to the formation of the Dirac cones below
$E_F$, for the $((3n+2)\times (3n+2))$ Ru periodicity, whereas in the 
($(3n+1)\times (3n+1)$) Ru/graphene systems, the barycenter C atom contribute
to the formation of the Dirac cones above $E_F$. 
In the region of intergration used to calculate the LDOS around the C2
cone (formed by bands with up spin), there are states with opposite spin, which
are associated with the formation of the cone C3, as shown in Fig. \ref{Fig3} (b).
These states, although located around 0.5~eV above the vertex of the cone C3, have a 
distribution of peaks in the LDOS [ shown in Fig. \ref{Fig4} (a-right)] similar to
that presented at the vertex of the cone C3 [ shown in \ref{Fig4} (b-right)].
The same behavior occurs for the other Dirac cones (C1-C4). 
Thus, the pattern of peaks distribution is not a characteristic only of the
vertex of the cones, but a characteristic of the entire energy band that forms
the cones.

In contrast, the band structures of the $(3n\times3n)$
Ru/graphene systems do not show Dirac cones. In this case, the $K$ and
$K^\prime$ points are folded into the $\Gamma$ point, and upon the presence of
Ru adatoms in a $(3n\times3n)$ periodicity, those electronic states face an
intervalley scattering process suppressing the formation of the Dirac
cones\cite{Ding2011}. Figure~5(a) presents the electronic band structure of a
(6$\times$6) Ru/graphene system, where we find an energy gap of $0.11$~eV at the
$\Gamma$ point. In this case, we find a quite different LDOS distribution
[Fig~5(b)] on the graphene sheet, when compared with the other
$((3n+1)\times(3n+1))$ and $((3n+2)\times(3n+2))$ Ru/graphene systems. Namely,
the highest occupied (spin-up) states, at the $\Gamma$ point, spread out
somewhat homogeneously on the graphene sheet. Similar results (not shown) are
found for the lowest
unoccupied (spin-down) states, as well as for other $(3n\times3n)$
periodicities. We find no LDOS peaks in the hexagonal lattice of barycenters,
which can be attributed to the absence of carbon atoms, since for such Ru
periodicity the barycenters lie at the hollow site of the graphene sheet. 

\begin{figure}
\includegraphics[width = 8.5cm ]{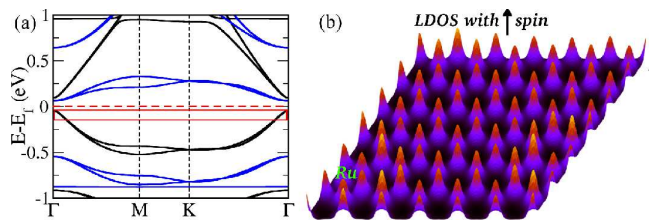}
\caption{(Color online) (a)Electronic band structure and (b) LDOS for the
(6$\times$6) Ru/graphene system. The region of integration used to make the LDOS
is represented by the (red) box in (a).}
\label{Fig5}
\end{figure} 

In order to provide further support to such subtle compromise between the
arrangement of the Ru adatoms on the graphene sheet and the formation of the
Dirac cones, we have examined two additional configurations for Ru adatoms on
graphene. That is,  Ru adatoms forming  rectangular and hexagonal lattices. For
those Ru/graphene systems, we find two (spin-polarized) Dirac cones, instead of
the four ones present at the triangular deposition of Ru on graphene. The
disappearance of two cones occurs since the surface potential induced by the Ru
adatoms has no longer a hexagonal symmetry. 
Here, we conclude that the presence of four Dirac cones is constrained by the 
triangular arrangement of the Ru adatoms on the graphene host.

\begin{figure*}
\centering
\includegraphics[width = 17 cm ]{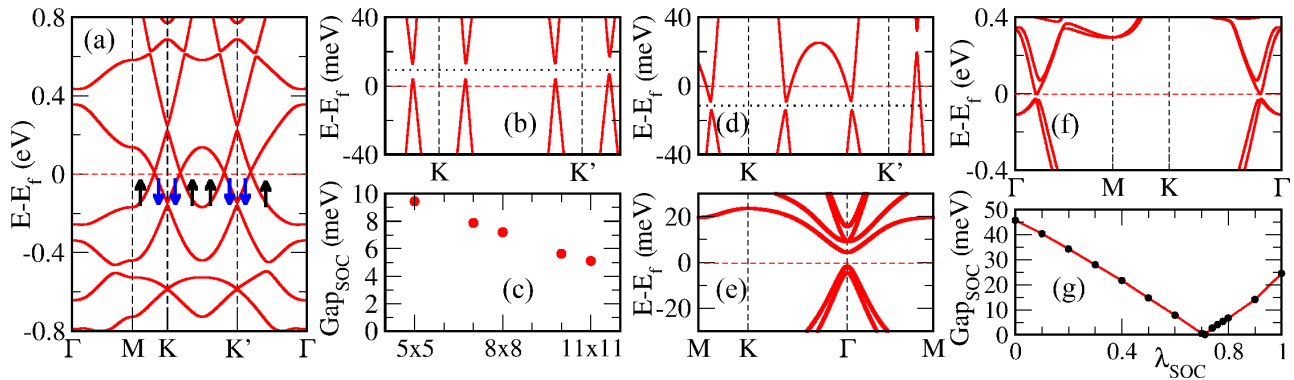}
\caption{(Color online) The SOC effects in Ru/graphene systems. (a) Electronic band structure for the 
(5$\times$5) periodicity with SOC. (b) Band structure of
(a) near the Fermi level. (c) Energy band gaps for the
$((3n+1)\times(3n+1))$ and $((3n+2)\times(3n+2))$ families with periodicities
$5\times5$ or larger. 
Band structures for the (d) (4$\times$4), (e) (6$\times$6) and (f) (2$\times$2) periodicities. 
(g) Band gap variation of (f) with the SOC strength $\lambda_{SOC}$. 
The dashed (red) lines indicate the Fermi levels, whereas the black dotted lines
is just to better visualize the energy gap. }
\label{Fig7}
\end{figure*}

By turning on the SOC we have the ingredients necessary to look for topological
phases in Ru/graphene systems. 
Since in pristine graphene the radial contribution of the SOC term is
negligible, and both the spin and the orbital magnetic moments are quenched, the
properties of Ru/graphene system are defined by the 4d Ru orbitals. In the energy
range in which the Dirac cones C1-C4 are formed, there are similar contributions
coming from the C 2p ($\pi$) and the Ru 4d orbitals, as already discussed above. This Ru
contribution will give rise to the SOC effects. We analysed the wave functions
around the Fermi energy, and concluded that their coefficients present significant
changes only close to band crossings. Thus, the SOC does not modify either the
electronic configuration or the effective population of the 4d orbitals when compared to the case when only the spin polarization is included.
Likewise, with the SOC the total energy of the system decreases only by 0.48meV,
so that the change in binding energy is negligible. The most relevant band
crossings occur at the Dirac cones above and below the Fermi energy as well as right
at the Fermi energy. The SOC will open gaps at these band crossings.

In order to understand how the inclusion of this interaction contributes
to the formation of energy gaps, we separately studied the diagonal and off-diagonal
contributions of $\boldsymbol{L}\cdot\boldsymbol{S}$ to $\boldsymbol{V}^{SOC}$
[see eqs. (\ref{eqls1}) and (\ref{eqls2})].

We find that the off-diagonal term breaks the
degeneracies at the $K$ point, opening gaps at the Dirac cones C1-C4.
Without the SOC, the Dirac cones are formed by intersections of bands with the same spin, where
the states have a unique non null component of the spinor.
As previously discussed, these bands have contributions from the orbital angular momentum $l=2$ ($4d$ orbitals) with
$m=\pm1$ and $\pm2$ ($d_{x^{2}-y^{2}}$, $d_{xz}$, $d_{yz}$ and $d_{yx}$
orbitals), leading to non-null off-diagonal terms. 
Through the self-consistent-cycle, these off-diagonal terms generate wavefunctions
with non null coefficients at the spinor component which was previously zero,
and break the degeneracies at the Dirac cones, opening an energy gap. 


We also find that the diagonal elements contribute to the opening
of gaps at the Fermi level, in the vicinity of $K$ and $K'$ points (see Fig. 6(a)).
In this case, there are two effects:
(i) The exchange and correlation potential generates a non-collinear spin
coupling term via $V^{xc,\sigma-\sigma}$; and 
(ii) the splitting of the energy bands with opposite spins is generated by
the addition (subtraction) of the matrix element $\langle l_{i},m_{i}|L_{z}|l_{j},m_{j}\rangle$
in $H_{ij}^{\uparrow\uparrow}$ ($H_{ij}^{\downarrow\downarrow}$). 
The addition of these two effects leads to the opening of gaps at the Fermi
energy.

In Fig. \ref{Fig7} (a) we present the electronic band structure of the (5$\times$5)
Ru/graphene system. The SOC gives rise to energy gaps of 9.5~meV right at the
points where the different spin band cross.
characteristic of the QAH phase\cite{Weeks2011} as depicted in Fig. \ref{Fig7} (b). 
In Fig. \ref{Fig7} (a) the spin texture is indicated, and is characteristic of a
QAH topological phase.
Also, we calculated the Chen number with the Eq. \ref{eq1}, obtaining
$\mathcal{C} = -2$ for the (5$\times$5) Ru/graphene, unequivocally confirming the
QAH phase.

As discussed above, the electronic properties of Ru/graphene system,
such as  the strength of the exchange field ($E_{x}$), the energy separation between the
Dirac cones ($\Delta_{C2-C3}$), and the electronic contribution of Ru $4d$  to
the formation of the Dirac cones $C1-C4$, all 
depend on the Ru concentration and ($n \times n$) periodicity. 
In this work we also found an intriguing dependence between the topological
phase and the ($n \times n$) periodicity of Ru/graphene. 
Indeed, by calculating the Chern number we found $\mathcal{C} = -2$ for all 
$((3n+1)\times(3n+1))$ and $((3n+2)\times(3n+2))$ Ru/graphene systems with
periodicities ($5\times5$) or larger. For those systems, the non-trivial 
band gap vary with the periodicity as shown in Fig \ref{Fig7} (c).
On the other hand, for the ($4\times4$) Ru/graphene system, the
Ru$\leftrightarrow$Ru interaction is strengthened, and the
Ru $4d$ orbitals become less localized. 
For this periodicity, with the SOC turned off the crossings between the up and
down bands are not all aligned in energy, and with the SOC turned on  
the opening of gaps occurs at different energies, leading to a non-gapped band
structure (metallic states), as shown in Fig. \ref{Fig7} (d) . These 
metallic states prevent the observation of the QAH effect. However, 
we obtain a non-null Chern number ($\mathcal{C}\approx 0.98$), indicating a finite Anomalous Hall
Conductivity, which is given by: $\sigma_{xy}=\mathcal{C}\frac{e^{2}}{\hbar}$.
For all the  ($3n\times3n$) Ru/graphene
systems we find $\mathcal{C} = 0$. This is a consequence of the trivial bandgap
at the $\Gamma$ point, generated by the intervalley $K$ and $K'$ scattering
process [the band is shown in Fig. \ref{Fig7} (e)]. Thus, all the
($3n\times3n$) Ru/graphene systems are trivial insulators since the SOC is not
strong enough to reverse the trivial band gap.
Further increase on the Ru
concentration (0.25 ML of Ru adatoms) was examined by considering the
(2$\times$2) periodicity.
In these systems, the barycenters are located at the NNN C sites to the Ru adatoms. 
The $(2\times 2)$ Ru/graphene exhibits
quite different electronic and topological properties in comparison with 
the other Ru/graphene systems, because: 
(i) it presents an energy band gap, and zero net magnetic moment [see Fig.
\ref{Fig7} (f)];
(ii) it presents the QSH phase.
Here, we use the adiabatic continuity argument to prove
(ii). This argument has been used to identify
2D and 3D topological insulators\cite{PhysRevLett.105.036404,PhysRevLett.106.016402,bernevig2006quantum,PhysRevB.76.045302,PhysRevB.78.045426,PhysRevB.80.085307,padilha2013quantum}.
According to this argument, if
the Hamiltonian of a system is adiabatically transformed into another, the
topological classification of the two systems can only change if the band gap
closes.
Thus, we smoothly changed the SOC strength by placing a multiplicative
factor, $\lambda_{SOC}$, in the term associated with the on-site approximation
of the SOC Hamiltonian, $H_{SOC}$. When this parameter is varied from zero to
one, we observed a variation of the gap as shown in Fig.~ \ref{Fig7}(g).
At $\lambda_{SOC}=0.712$ we find a metallic state (gap closing), generated by an
inversion of the states that contribute to the formation of the HOMO and LUMO,
indicating a transition from one trivial topological state (without SOC,
$\lambda_{SOC}=0$) to the QSH state (with SOC, $\lambda_{SOC}=1$). 

The above description is sumarized in Table \ref{tab1}, where can be
apreciated the multiple topological phases exhibited by the Ru/graphene systems.

\begin{table}
\caption{\label{tab1} Multiple topological phases in Ru/graphene systems as a function of
the periodicity.}
\begin{ruledtabular}
\begin{tabular}{cc}
 periodicity & topological phase \\ \hline
 ($2\times2$) & QSH \\
 ($4\times4$) & metal \\
 $((3n+1)\times(3n+1))$\footnotemark[1] & QAH \\
 $((3n+2)\times(3n+2))$\footnotemark[2] & QAH \\
 $(3n\times3n)$\footnotemark[2]         & trivial insulator 
\footnotetext[1]{for $n\ge 2$}
\footnotetext[2]{for $n\ge 1$}
\end{tabular}
\end{ruledtabular}
\end{table}

\section{Summary}

In summary, based on {\it ab initio} calculations, we investigate the
structural and the electronic properties of graphene adsorbed by Ru adatoms,
Ru/graphene. We  map the evolution of the electronic charge density distribution
around the Fermi level as a function of different Ru/graphene
geometries. We found that the Ru adatom fixes the wave function phase of its NN and
NNN C atoms, whereas the Ru$\leftrightarrow$Ru interaction, mediated by the $\pi$
orbitals of the graphene sheet, gives rise to four spin-polarized Dirac cones
for the $((3n+1)\times(3n+1))$ and $((3n+2)\times(3n+2))$ Ru/graphene systems.
The  electronic distributions of the states that form those Dirac cones
are constrained by the periodicity of the Ru adatoms on the graphene host. 
For triangular arrays of Ru adatoms, four spin-polarized Dirac cones are generated by a suitable 
coupling between the electronic states of two hexagonal
lattices, one composed by the carbon atoms of the graphene host, and the other
attributed to the (barycenter) surface potential on the graphene sheet induced
by the triangular lattice of Ru adatoms.  
 For other geometries, hexagonal and rectangular, we have only two
spin-polarized Dirac cones, while there are no Dirac cones for  $(3n \times 3n)$
Ru/graphene. The inclusion of SOC promotes multiple topological phases when 
graphene is doped with triangular arrays of Ru.
The topological phase in those systems depends on the periodicities (or concentration) of Ru adatoms on the 
graphene sheet. For a high coverage in the $(2\times2)$ periodicity (25\%) of Ru adatoms the QSH phase is
present, whereas for the $((3n+1)\times(3n+1))$ and $((3n+2)\times(3n+2))$ Ru/graphene 
systems the QAH phase will be preserved even for low coverage of Ru adatoms 
(less than 1\%). These results are summarized in the Table \ref{tab1}.

Even though transition metals adatoms have been used before to obtain distinct
non-trivial topological phases in graphene, in previous works it was always
considered that distinct transition metals would provide distinct topological
phases. However, we have shown that this is not necessarily so. The same
transition metal can provide distinct topological phases, depending on the
particular geometrical arrangement.


\section{Acknowledgements}
The authors would like to thank Prof. Shengbai Zhang for fruitful discussions.
Also, we would like to thank the financial support by Conselho Nacional de Desenvolvimento Cient\'ifico e
Tecnol\'ogico/Institutos Nacionais de Ci\^encia e Tecnologia do Brasil
(CNPq/INCT), the Coordena{\c c}\~ao de Aperfei{\c c}oamento de Pessoal de
N\'ivel Superior (CAPES), and the Funda{\c c}\~ao de Amparo \`a Pesquisa do Estado de
S\~ao Paulo (FAPESP).

\bibliography{Ref} 


\end{document}